\documentclass[
	reprint,
	aps,
	prl,
	amsmath,amssymb,
	superscriptaddress,
	floatfix
]{revtex4-2}

\usepackage{graphicx}
\usepackage{dcolumn}
\usepackage{booktabs}
\usepackage{physics}
\usepackage{bm}
\usepackage{hyperref}
\usepackage[capitalize]{cleveref}

\newcommand\Qph{Q_\text{ph}}
\newcommand\Pph{\Pi_\text{ph}}
\newcommand\vlambda{\boldsymbol{\lambda}}
\newcommand\veps{\boldsymbol{\epsilon}}
\newcommand\Omegar{\tilde{\Omega}}
\newcommand\ms{m_\text{S}}
\newcommand\Hn[1]{H\textsubscript{#1}}
\newcommand\Dnh[1]{D\textsubscript{#1h}}
\newcommand\CBD{C\textsubscript{4}H\textsubscript{4}}

\begin{document}
\title{Engineering molecular potential energy surfaces \\ using magnetic cavity quantum electrodynamics}

\author{Lukas Weber}
\affiliation{Center for Computational Quantum Physics, The Flatiron Institute, 162 Fifth Avenue, New York, NY 10010, USA}
\author{Leonardo dos Anjos Cunha}
\affiliation{Center for Computational Quantum Physics, The Flatiron Institute, 162 Fifth Avenue, New York, NY 10010, USA}
\author{Johannes Flick}
\affiliation{Center for Computational Quantum Physics, The Flatiron Institute, 162 Fifth Avenue, New York, NY 10010, USA}
\affiliation{Department of Physics, City College of New York, New York, New York 10031, United States}
\affiliation{Physics Program, Graduate Center, City University of New York, New York, New York 10016, United States}
\author{Shiwei Zhang}
\affiliation{Center for Computational Quantum Physics, The Flatiron Institute, 162 Fifth Avenue, New York, NY 10010, USA}

\begin{abstract}
We investigate the effects of coupling a quantum-magnetic cavity field to molecules.
Our high-precision auxiliary-field quantum Monte Carlo 
calculations capture the effect of the cavity field in the 
presence of electron correlations, and their interplay and competition.
In \Hn{2}, we find that a strong enough cavity coupling makes the original bound ground state metastable, along with inverting the singlet-triplet gap. In ring molecules (e.g., \Hn{$n$}), the magnetic cavity coupling stabilizes symmetric geometries. As a consequence, open-shell rings such as \Hn{4}, \Hn{8}, or \CBD{}, which would undergo Jahn-Teller distortions outside of the cavity, obtain exotic spin or ring-current polarized, antiaromatic ground states. These effects are enhanced by increasing the molecule concentration inside the cavity. Our results suggest cavity quantum electrodynamics beyond the long-wavelength approximation as a promising avenue for cavity-altered chemistry.
\end{abstract}

\maketitle
\section{Introduction}
Within the last years, the idea of coupling cavity quantum electrodynamic (QED) fluctuations to materials or molecules~\cite{ruggenthaler_quantumelectrodynamical_2014,hubener_engineering_2021,schlawin_cavity_2022,hubener_quantum_2024} to alter their properties has seen a bloom of promising experimental results, showing various cavity-modified quantum phases \cite{appugliese_breakdown_2022,jarc_cavitymediated_2023,enkner_tunable_2025,keren_cavityaltered_2026} as well as chemical reactions \cite{thomas_groundstate_2016,thomas_tilting_2019,vergauwe_modification_2019,hirai_modulation_2020,thomas_ground_2020,lather_cavity_2021,lather_improving_2021,sau_modifying_2021,ahn_modification_2023}. While a comprehensive understanding is still in progress, in quantum chemistry, the input from these experiments has giving rise to several theories based on cavity-modified vibrational energy transfer~\cite{li_cavity_2021,Schaefer_2022,lindoy_quantum_2023,du_vibropolaritonic_2023,vega_theoretical_2025} or spin glass analogies \cite{sidler_perspective_2022,sidler_unraveling_2024,sidler_collectivelymodified_2026} that address resonance behaviors in the cavity frequency that earlier approaches failed to account for.

Most of these prior experiments and theoretical works, however, consider cavities where the coupling to matter is predominantly electrical. While such electric cavities experimentally achieve strong coupling to the electronic and nuclear charges of the matter, they are also subject to “no-go” theorems~\cite{rzazewski_phase_1975,bialynicki-birula_nogo_1979,gawedzki_nogo_1981,andolina_cavity_2019} that prevent the ground-state condensation of cavity photons---also known as the superradiant quantum phase transition~\cite{hepp_superradiant_1973}. As a side effect, all ground-state effects of the cavity are relegated to second- or higher-order field-fluctuation processes. This can be seen as one of the reasons why modeling and understanding cavity experiments has proved challenging, requiring sophisticated models that take higher-order processes into account~\cite{tasci_photon_2025,tasci_super_2025}.

Magnetic cavities, on the other hand, do not suffer from this limitation: they display ground-state photon condensation that is similar to the superradiant quantum phase transition, but in fact equivalent to cavity-induced magnetism~\cite{andolina_theory_2020,guerci_superradiant_2020} and thus straightforwardly a physical effect. This insight has motivated several theoretical proposals for magnetic cavities coupled to solids, either to observe their photon condensation~\cite{roman-roche_photon_2021,andolina_can_2022,bacciconi_firstorder_2023} or topological phases of matter~\cite{guerci_superradiant_2020}. The effect of magnetic cavities has been recently investigated in molecular systems \cite{fischer_cavitymodified_2025,fischer_spincavity_2026}. In these works, the authors employ model Hamiltonians for molecular systems to investigate cavity modified Zeeman and SOC effects. Meanwhile, the ultrastrong coupling regime has already been achieved between magnons and photons in magnetic cavities~\cite{goryachev_highcooperativity_2014,zhang_ultrastrong_2019,golovchanskiy_approaching_2021,macedo_map_2025,zhang_simultaneous_2025}, putting the exploitation of magnetic cavity coupling for controlling materials properties within reach of the next generation of experiments.

In this work, we show how magnetic cavity coupling can be used to engineer molecular potential energy surfaces. 
Molecular systems provide an ideal platform. 
They are pristine, where different effects can be disentangled and investigated more readily. 
High-precision calculations can be performed, especially with 
the recently developed high-precision QED auxiliary-field quantum Monte Carlo (QED-AFQMC)~\cite{weber_phaseless_2025,weber_lightmatter_2025} method, to make reliable predictions based on the \emph{ab initio} Hamiltonian. 
These systems also point to a realistic route for experimental realization with, as our results indicate, 
a strong likelihood to achieve observable effects.
We find that in \Hn{2} molecules, coupling to a magnetic cavity can lead to an inversion of the singlet-triplet gap and make the bound singlet ground state metastable. We further show that in hydrogen rings (\Hn{$n$}) and cyclobutadiene (\CBD), the magnetic cavity coupling favors symmetric states, which can stabilize antiaromatic exotic spin or current polarized ground states that, without the cavity, would be prevented by Jahn-Teller distortion~\cite{pritchard_new_2019}. These effects are further enhanced by increasing the molecule concentration. Our results are obtained using 
QED-AFQMC,
supplemented by full configuration interaction (QED-FCI)~\cite{haugland_coupled_2020}, and unrestricted Hartree-Fock (QED-UHF)~\cite{haugland_coupled_2020} calculations. All methods are modified to work beyond the dipole approximation, as required for simulating magnetic fields.

\section{Results}
In the following, after a brief description of the model, we present our numerical results for the magnetic cavity coupled \Hn{2} molecule. Afterward, we focus on two kinds of ring molecules: the hydrogen rings H$_n$ with $n=2,4,6,8$ and fixed radius $R$, as well as cyclobutadiene \CBD. Finally, we comment on the concentration dependence of the observed effects.

\subsection{Magnetic cavity model}
We describe the magnetic cavity coupled system using a Pauli-Fierz Hamiltonian in the Coulomb gauge, which in atomic units reads
\begin{align}
H &= H_\text{mol} + \sum_i \qty[ -\vu{A}(\vb{r}_i) \cdot \vb{p}_i + \frac{\vu{A}^2(\vb{r}_i)}{2}] \nonumber\\
&- \sum_i \frac{g_e}{2} \vb{S}_i \cdot \qty(\nabla\times \vu{A}(\vb{r}_i)) + \frac{\Omega}{2} (\Pph^2 + \Qph^2 - 1),\label{eq:pfierz}
\end{align}
with the regular matter Hamiltonian $H_\text{mol}$ containing electron kinetic energy, as well as electron-electron, electron-nuclear, and nuclear-nuclear interactions.
$\vb{S}_i$, $\vb{p}_i$, and $\vb{r}_i$ are the spin operator,
momentum, and position, respectively, of the $i$th electron, $\vu{A}(\vb{r})$ the quantum vector potential and $\Qph$ ($\Pph$) is the photon displacement (momentum). We set $g_e = 2$ in the following and treat the nuclei as quasistatic within the Born-Oppenheimer approximation so that their contribution to the magnetic interaction vanishes.

To describe a magnetic cavity field, it is necessary to break the long-wavelength approximation and make $\vu{A}(\vb r)$ position dependent. Assuming the coupling to the cavity is dominated by a single mode of frequency $\Omega$ with approximately constant magnetic field in the interacting region, we set $\vu{A}(\vb r) = \Qph \vlambda \times \vb r /\sqrt{\Omega}$. Apart from a spatially constant magnetic field $\vu{B}=2\Qph \vlambda/\sqrt{\Omega}$, this mode also gives rise to a vortex quantum electric field that still breaks translation invariance. A detailed discussion of the resulting origin dependence can be found in Supplementary Note~1. We finally rewrite \cref{eq:pfierz} in the standard Zeeman form:
\begin{align}
H &= H_\text{mol} + \frac{\Omega}{2} (\Pph^2 + \Qph^2 - 1) \nonumber\\
&+ \sum_i \qty[ \vlambda^2 \vb{r}_i^2 - (\vlambda \cdot \vb{r}_i)^2] \frac{\Qph^2}{2\Omega}\nonumber\\
&- \sum_i \frac{1}{2} (\vb{L}_i + 2\vb{S}_i) \cdot \vu{B}. \label{eq:ham}
\end{align}
To facilitate a coupling between the cavity magnetic field and electronic currents, we consider a cavity frequency on the order of charge excitations, $\Omega = 0.05$ Ha, throughout this work. A more systematic study of the frequency dependence is left to future work. We adopt a coordinate frame where $\vlambda = \lambda \mathbf{e}_z$. All calculations are performed in augmented correlation-consistent Gaussian basis sets~\cite{kendall_electron_1992,peterson_benchmark_1994,pritchard_new_2019}.

\subsection{\Hn{2}}

\begin{figure}
\includegraphics{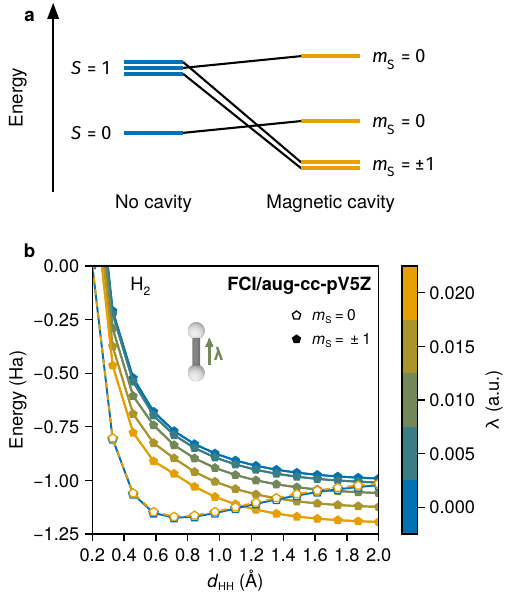}
\caption{\textbf{\Hn{2} coupled to a magnetic cavity.} (a) Sketch of the lowest energy levels in cavity-coupled \Hn{2}. Without light-matter coupling, there is a singlet ($S=0$) bonding ground state and a triplet ($S=1$) antibonding excited state. Under finite cavity coupling, through the quantum spin Zeeman effect, the triplet splits into an $\ms=0$ singlet and an $\ms=\pm 1$ doublet of lower energy. The $\ms = 0$ states get shifted upwards weakly by the diamagnetic coupling. (b) Potential energy surfaces as a function of the bond length $d_\mathrm{HH}$ and the cavity coupling $\lambda$, calculated using FCI in the aug-cc-pV5Z basis set. For each coupling, only the lowest energy $\ms = 0, \pm 1$ states are shown. For clarity, in the $\ms = 0$ surface, which depends only weakly on $\lambda$, the overlapping $\lambda=0.02$~a.u. curve is dashed. The molecule is aligned with the mode direction and centered at the origin.}
\label{fig:h2}
\end{figure}

We start the discussion of our numerical results with \Hn{2}, as the simplest molecule whose geometry can be influenced by a magnetic cavity. Outside of the cavity, \Hn{2} has a singlet  ($S=0$) bonding energy surface and a triplet ($S=1$) antibonding energy surface (\cref{fig:h2}(a)). The singlet energy surface is unaffected by the spin coupling term in \cref{eq:ham}, and the remaining orbital and diamagnetic contributions to the coupling are weak due to the geometry of \Hn{2}. The triplet energy surface, however, gets split: the magnetically polarized levels $S=1,~\ms = \pm 1$ both couple to the spin quantum Zeeman term, decreasing their energy. In contrast to the classical Zeeman effect, both magnetization directions get shifted down in energy by the same amount and remain degenerate.

Due to the small size of \Hn{2}, we can confirm this behavior 
with numerically exact QED-FCI calculations in a large basis set (\cref{fig:h2}(b)). 
This also provides an additional benchmark for our QED-AFQMC within the magnetic cavity QED setup here.  
These calculations show that there is a critical bond length beyond which the cavity coupled $\ms = \pm 1$ levels become the new ground state. This critical bond length decreases with increasing coupling, and at strong enough light-matter coupling, $\lambda \approx 0.02$~a.u., the original singlet bound state of \Hn{2} becomes metastable.

While for \Hn{2}, the effect of the cavity can be largely understood in terms of the quantum spin Zeeman effect alone, we will next shift our attention to ring molecules, where 
the orbital angular momentum around the ring and diamagnetic effects can also
play an important role.

\begin{figure*}[tp]
\includegraphics{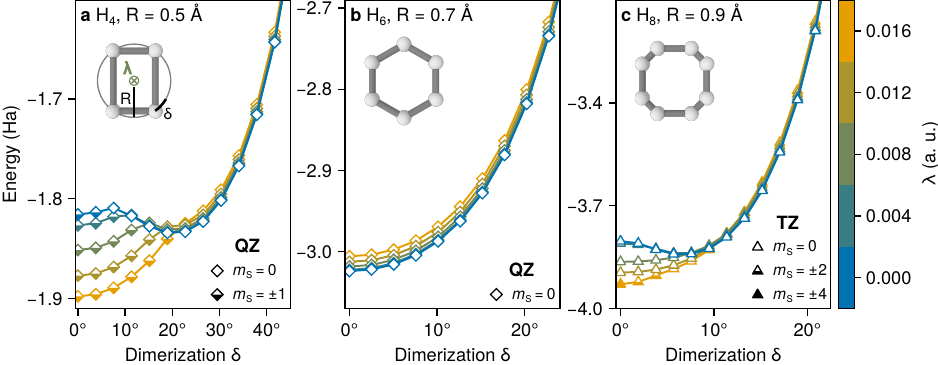}
\caption{\textbf{Potential energy surfaces of cavity-coupled hydrogen rings.} For each dimerization angle $\delta$ and magnetic cavity coupling $\lambda$, only the lowest-energy spin state, $\ms$, is shown. Calculations are performed using QED-UHF at a fixed ring radius $R$ in the aug-cc-pVTZ (TZ) and in aug-cc-pVQZ (QZ) basis sets. The insets show a sketch of the ground state ring geometries outside of the cavity. The center of the ring is located at the origin, and the quantum magnetic field is perpendicular to the ring plane.}
\label{fig:hrings}
\end{figure*}
\subsection{H rings}
As a proof-of-principle model for a ring molecule, we first consider the hydrogen rings H$_n$ with a fixed radius $R$. Under this constraint, the arrangement of the H nuclei on the ring can be subject to the Jahn-Teller effect, e.g.~via the Peierls instability, which leads to dimerized geometries~\cite{peierls_quantum_1995,frohlich_theory_1954,longuet-higgins_alternation_1959}. To investigate this instability, we parameterize the geometry of the ring as $\vb{r}_{k} = R\, (\cos \alpha_k, \sin \alpha_k, 0)$ with $\alpha_{k} = 2\pi k/n + (-1)^k \delta$ and a dimerization angle $\delta$~\footnote{We note that, besides the Peierls distortion, the H rings could also display other highly asymmetric distortions, as has been shown for $4n$-site Hubbard rings~\cite{lieb_stability_1995}. For the sake of studying a model system, we focus on the Peierls distortion as a representative example.}. The radius $R$ is scaled with $n$ so that the average hydrogen-hydrogen distance is approximately 0.7 \AA, slightly below the equilibrium bond length of \Hn{2}.

\Cref{fig:hrings} shows the potential energy surfaces resulting from a QED-UHF calculation for $n = 4, 6, 8$ for different coupling strengths $\lambda$. 
We perform these UHF calculations to obtain  an
overview by scanning many parameters relatively quickly, but also to benchmark them by comparison with our AFQMC calculations discussed below.
At the investigated radii, outside the cavity ($\lambda = 0$), the energy of \Hn{4} and \Hn{8} is minimized by a dimerized configuration, $\delta \neq 0$, while \Hn{6} has a symmetric ground state.

This situation changes at finite cavity couplings. Both \Hn{4} and \Hn{8} develop a minimum at $\delta=0$ around $\lambda \approx 0.008$~a.u. For \Hn{4}, this minimum is spin-polarized in the z direction with $\ms = \pm 1$ (which, interestingly, is already present outside of the cavity as a metastable state). For \Hn{8}, by contrast, there is an intermediate regime where the symmetric minimum is spin unpolarized. To obtain a downshift of an spin unpolarized state at the QED-UHF level, this regime must instead be loop-current polarized and couple to the orbital quantum Zeeman term. \Hn{6} always retains a symmetric ground state geometry.

For all three rings, we observe that for very large couplings $\lambda$, the aug-cc-pVTZ and aug-cc-pVQZ basis sets can no longer accurately describe the ground state. We therefore limit our discussion to $\lambda \le 0.016$~a.u., where this problem does not appear, and we find our calculations to be sufficiently converged. Supplementary Note 1 contains a detailed discussion of the basis set convergence, including an investigation of the origin dependence of our calculations.

\begin{figure}
\includegraphics{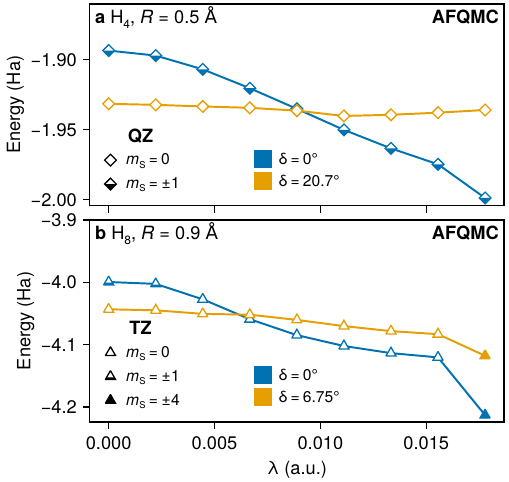}
\caption{\textbf{Stability of different hydrogen ring geometries under magnetic cavity coupling.} Energies as a function of the magnetic cavity coupling $\lambda$, for the symmetric geometry ($\delta = 0$) and the optimal QED-UHF geometry outside of the cavity ($\delta \ne 0$), calculated using QED-AFQMC in the aug-cc-pVTZ (TZ) and aug-cc-pVQZ (QZ) basis sets. For each dimerization angle $\delta$ and coupling, only the lowest-energy spin state, $\ms$, is shown. The statistical error bars are smaller than the markers.}
\label{fig:hrings_qmc}
\end{figure}
To confirm that the cavity-induced stabilization of symmetric ground states is not an artifact of the lack of electron-electron and electron-photon correlation effects in the QED-UHF ansatz, we perform QED-AFQMC calculations of \cref{eq:ham}, with results shown in \cref{fig:hrings_qmc}. In both \Hn{4} and \Hn{8}, these calculations confirm the presence of a cavity-stabilized symmetric ground state. In \Hn{4} (\cref{fig:hrings_qmc}(a)), the transition happens at $\lambda = 0.0091(1)$~a.u., while for \Hn{8} (\cref{fig:hrings_qmc}(b)), it is shifted down to $\lambda \approx 0.0061(1)$~a.u. A further correlation effect in \Hn{8} is the stabilization of $\ms = \pm 1$ states for $\delta = 0$ at small $\lambda$.

Our results for H rings show that both the spin and orbital angular momentum degrees of freedom can play a crucial role in magnetic cavity coupling, depending on the molecular geometry. Furthermore, more dramatic effects are observed in open-shell systems that are already unstable outside of the cavity. However, as we artificially fix their radius, the H rings should be considered model systems. In the next section, we show that the same effect can also be observed in a realistic carbon ring molecule, cyclobutadiene.

\subsection{Cyclobutadiene}
\begin{figure}
\includegraphics{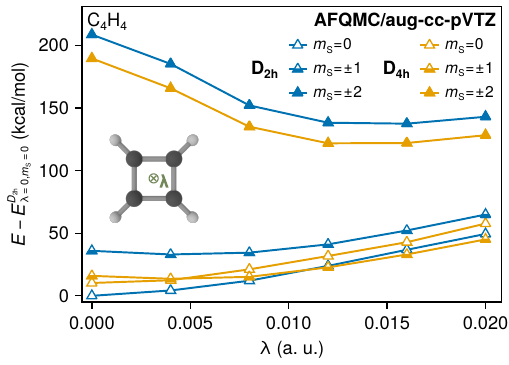}
\caption{\textbf{Stability of different cyclobutadiene geometries under magnetic cavity coupling.} Energies as a function of the magnetic cavity coupling $\lambda$. The \Dnh{2} and \Dnh{4} geometries are fixed to their $\lambda=0$ values (\cref{tab:geometry}). The statistical error bars are smaller than the markers.}
\label{fig:cyclobutadiene}
\end{figure}
\begin{table}
\caption{\textbf{Geometries of cyclobutadiene simulations.} Cavity-relaxed carbon-carbon distances are obtained by fitting paraboloids to the minima in \cref{fig:cyclobutadiene_perturbation}. The parameters obtained from AFQMC are given with one standard deviation statistical uncertainties. All lengths are in angstrom.}
\begin{ruledtabular}
\label{tab:geometry}
\begin{tabular}{lcccc}
\textbf{No cavity} & \multicolumn{4}{l}{CASPT2(12,12)/cc-pVTZ\footnote{from Ref. \cite{monino_reference_2022}}}\\
Symmetry group & $d_\text{CC,1}$ & $d_\text{CC,2}$ & $d_\text{CH}$ & $\angle$H--C=C \\
\midrule
$D_\text{2h}$ & 1.354 & 1.566 & 1.077 & 134.99° \\
$D_\text{4h}$ & 1.449 & 1.449 & 1.076 & 135.00° \\
\midrule
\textbf{Cavity} & \multicolumn{4}{l}{AFQMC/aug-cc-pVTZ}\\
Symmetry group & $d_\text{CC,1}$ & $d_\text{CC,2}$ & $d_\text{CH}$ & $\angle$H--C=C \\
\midrule
$D_\text{2h}$ & 1.344(2) & 1.538(1) & [1.077]\footnotemark[2] & [134.99°]\footnotemark[2] \\
$D_\text{4h}$ & 1.4269(5) & 1.4269(5) & [1.076]\footnotemark[2] & [135.00°]\footnotemark[2] \\
\end{tabular}
\end{ruledtabular}
\footnotetext[2]{taken from outside of the cavity.}
\end{table}
\begin{figure}
\includegraphics{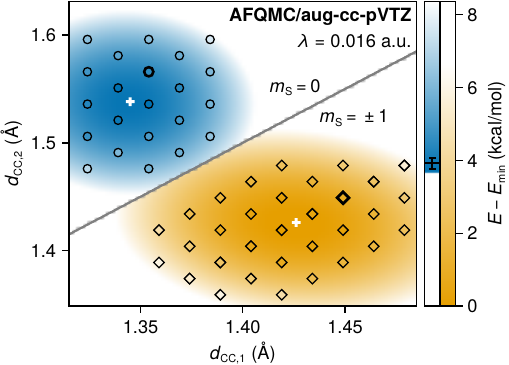}
\caption{\textbf{Potential energy surface of cyclobutadiene in the cavity.} Perturbations of the two carbon-carbon
bond lengths,
$d_\text{CC,1}$ and $d_\text{CC,2}$, around the D\textsubscript{4h}/$\ms=\pm 1$ (yellow) and D\textsubscript{2h}/$\ms = 0$ (blue) equilibrium geometries.
The out-of-cavity equilibrium carbon-carbon bond lengths are indicated by markers with thick outlines.
The hydrogen-carbon bond lengths and angles are left constant. The color of the markers corresponds to the QED-AFQMC results in the aug-cc-pVTZ basis set. The colors in the background show a parabolic fit. The center of the fit is marked in white.  
On the left colorbar, the gap between the fitted energy minima is marked along with its statistical error.
The magnetic cavity coupling is fixed at $\lambda = 0.016$\,a.u.}
\label{fig:cyclobutadiene_perturbation}
\end{figure}
In this section, we consider magnetic-cavity-coupled cyclobutadiene, \CBD. As with the hydrogen rings, we consider planar geometries perpendicular to the cavity polarization $\vlambda$. UHF calculations incorrectly predict that cyclobutadiene has a symmetric ground state outside of the cavity. Our analysis below are all based on QED-AFQMC results.

Starting with fixed geometries from the literature~\cite{monino_reference_2022} for the most stable \Dnh{2} and \Dnh{4} configurations outside the cavity, we perform a scan in $\lambda$, with results presented in \cref{fig:cyclobutadiene}. Similar to the case of the H rings, at $\lambda = 0.0109(4)$ a.u., there is a level crossing to the more symmetric \Dnh{4} geometry with finite magnetization $\ms=\pm1$. In contrast to the H rings, highly polarized states with $|\ms| \ge 2$ stay energetically well separated from the ground state even for large $\lambda$. This can be attributed to a stronger diamagnetic response due to the larger density of electrons. 

Outside of the cavity, QED-AFQMC predicts an energy difference $E^\text{\Dnh{4}}_{\ms=0}-E^\text{\Dnh{2}}_{\ms=0} = 10.3(5)\,\text{kcal/mol}$, slightly more than the results from Ref.~\cite{monino_reference_2022}, whereas for $\lambda=0.016$ a.u., we find $E^\text{\Dnh{4}}_{\ms=0}-E^\text{\Dnh{2}}_{\ms=\pm 1} = -4.3(7)\,\text{kcal/mol}$. These results assume a fixed geometry from outside of the cavity.

To make more reliable statements, it is crucial to relax the geometry in the presence of light-matter coupling. To reduce the computational burden at the QED-AFQMC level, we assume that the hydrogen-carbon bond length and angle, which, out of the cavity, are nearly identical between $\Dnh{2}$ and $\Dnh{4}$ \CBD{}, also remain the same in the cavity, limiting our relaxation to the carbon bond lengths $d_\text{CC,1}$ and $d_\text{CC,2}$.

We find that the cavity-induced symmetric ground state is stable to these perturbations, as shown in \cref{fig:cyclobutadiene_perturbation}. Even though both potential energy minima relax to slightly smaller carbon-carbon distances, the global minimum retains \Dnh{4} symmetry. To quantify these findings, we fit paraboloids to the perturbed energies ($\chi^2_\text{\Dnh{2}}$/dof = 1.13, $\chi^2_\text{\Dnh{4}}$/dof = 0.55), and extract the relaxed carbon-carbon distances. 
The results are summarized in \cref{tab:geometry}, 
together with the out-of-cavity equilibrium bond lengths.
The resulting energies at the equilibrium geometry inside the cavity are $E^\text{min}_\text{\Dnh{2}} = -154.3957(2)$\,Ha and $E^\text{min}_\text{\Dnh{4}} = -154.4019(2)$\,Ha. Their energy difference is statistically compatible with that of the unrelaxed geometries.
\subsection{Enhancement with molecule concentration}
All of the preceding calculations are done for a single molecule in the cavity. For realistic experimental setups, it is crucial to consider the effect of a finite concentration of molecules interacting collectively with the cavity field. For H-ring molecules, we have shown that a mean-field decoupling of the cavity field can successfully describe the cavity-induced transitions in the ground-state geometry we identified. Based on a mean-field decoupling of the cavity, we will now show that the observed effects are enhanced by increasing the molecule concentration.

We start with \cref{eq:ham} and the assumption that 
the molecules are in the gas phase. 
We consider $N$ identical ring molecules that are far enough apart to not directly interact, but all coupled to the same cavity mode. In this approximation, i.e. a wavefunction of the form $\ket{\Psi} = [\bigotimes_{n=1}^N \ket{\psi_\text{mol}}] \otimes \ket{\chi_\text{ph}}$, the energy of the mean-field decoupled system becomes
\begin{align}
E_\text{MF} &= N E_\text{mol} + \frac{\Omega}{2} \langle \Pph^2 + \Qph^2 - 1\rangle \nonumber\\
&+ \expval{\mathbf{r}_\perp^2} \frac{N \vlambda^2 \langle\Qph^2\rangle}{2\Omega}\nonumber\\
&- N \expval{\vb{L} + 2\vb{S}} \cdot \frac{ \vlambda \expval{\Qph}}{\sqrt{\Omega}},
\end{align}
where $\vb{r}_\perp^2 = \vb r^2 - (\vlambda \cdot \vb r)^2/\vlambda^2$.
Here, we have used the fact that, at the mean-field level, it is possible to shift the origin of the vector potential for each molecule individually. Beyond the mean-field description, this is not possible, and there is an additional energy contribution which intrinsically depends on the absolute positions of the molecules in the cavity (see Supplementary Note 1).

At the minimum of $E_\text{MF}$, the cavity is in a squeezed coherent state, expressed in the eigenbasis of $\Qph$ as $\ket{\chi_\text{ph}} = (\frac{\pi}{s})^\frac14 \int \dd{q} e^{-\frac{s}{2} (q-q_0)^2} \ket{q}$, which we can use to express the energy per molecule in terms of purely matter quantities,
\begin{align}
\frac{E_\text{MF}}{N} &= E_\text{mol} + \frac{\Omegar}{2} - \frac{N}{\Omegar^2} \expval{(\mathbf{L} + 2\mathbf{S})\cdot \vlambda}^2,\label{eq:mft}
\end{align}
with the renormalized frequency $\Omegar^2 = \Omega^2 + N\vlambda^2 \expval{\mathbf{r}^2_\perp}$. This expression is equivalent to the single molecule problem with rescaled coupling $\vlambda \mapsto \sqrt{N}\vlambda$. Therefore, increasing the number of molecules in a constant mode volume is akin to increasing the coupling, while scaling the mode volume ($\vlambda \propto 1/\sqrt{V}$) at a constant concentration $N/V$ does not change the effect.

In an electrical cavity in the long-wavelength approximation, this same argument breaks down. Since in that case, ground states with $\expval{Q_\text{ph}} \neq 0$ are impossible, the current-current interaction term that would appear in the equivalent of \cref{eq:mft} always vanishes exactly. Therefore, the entire effect of the cavity is in the higher-order correlations beyond mean-field theory and has a weaker scaling with the number of particles. The collectivity we observe here is therefore a unique feature of working with magnetic cavities. For an ensemble of molecules, it can be viewed as a cavity-induced ferromagnetic transition.

In our current analysis, we have not considered different molecular orientations. Instead, we focused on the case in which the field is aligned with the rotation axis of the molecule. In a realistic setting, the molecules may start out randomly oriented. In cases of predominant $\mathbf{S}\cdot\hat{\mathbf{B}}$ coupling, such as in \Hn{2}, \Hn{4}, or \CBD{}, we anticipate the cavity-induced geometries to remain stable irrespective of the overall rotation. The predominant $\mathbf{L}\cdot\hat{\mathbf{B}}$ coupling in \Hn{8}, on the other hand, is directly related to the molecular orientation. Irrespective of changes to the geometry, the cavity should also induce collective molecular realignment. The direction of that alignment will likely depend on a competition between the diamagnetic response and the $\mathbf{L}\cdot\hat{\mathbf{B}}$ coupling as a function of the angle.

\section{Discussion and Outlook}
Using a combination of QED-UHF, QED-AFQMC, and QED-FCI simulations, we investigated the behavior of the H$_n$ and \CBD{} ring molecules strongly coupled to magnetic cavity modes. In all of these cases, we find that the coupling to the cavity prefers rotationally symmetric states or antibonding states in the case of \Hn{2}. The mechanism for this is the quantum electrodynamic equivalent of the Zeeman effect, which lowers the energy of magnetized or finite angular momentum states. If the cavity coupling is sufficiently strong, it can lead to changes in the ground-state magnetization, dissociation, or stabilization of antiaromatic geometries that would usually be precluded by Jahn-Teller distortions. We further find that these effects, which can also be interpreted as cavity-induced ferromagnetism, are enhanced by increasing the concentration of molecules in the cavity.

These findings stand in contrast to results reported for molecules coupled to quantum electrical cavity fields in two main ways. First, they do not require strong light-matter coupling at the individual level. Second, there is a clear intuition of what kinds of molecules and reactions are affected by magnetic cavities through the quantum Zeeman effect. This opens a large class of further candidate molecules that could be investigated for similar effects. In particular, it would be interesting to consider larger ring molecules with high magnetic susceptibility or the effect of magnetic solvents. Both of these may allow observing the geometry transitions reported here experimentally at even smaller coupling strengths. Based on the comparison of methods in our work, a mean-field treatment of the cavity should be a reasonable starting point for such investigations, provided that both orbital and spin magnetic couplings are included.

In a broader context, together with earlier proposals on altering material properties using magnetic cavities, our work highlights cavity coupling beyond the long-wavelength approximation as a promising approach for modifying matter in the strong light-matter coupling regime.

\section{Methods}
\subsection{QED-UHF}
Our UHF calculations are performed by numerically optimizing the variational energy of the product state $\ket{\text{SD}} \otimes \ket{\text{CS}}$ of an electronic Slater determinant and a photon coherent state. Resolving rotating electronic states at finite $\expval{\Qph}$ correctly requires the use of complex-valued orbitals. The optimization algorithm is LBFGS, implemented in Optim.jl~\cite{mogensen_optim_2018} with autodifferentiation from Zygote.jl~\cite{Zygote.jl-2018}. For both the H rings and cyclobutadiene, the combined light-matter energy landscape can be rugged. To ensure convergence to the global minimum, we start our optimizations from 100 random initializations.
\subsection{QED-AFQMC}
The AFQMC method \cite{zhang_constrained_1997} reformulates the imaginary time ground state projection of an interacting system as a branching random walk of many noninteracting systems. This random walk is then efficiently sampled using Monte Carlo methods. In the presence of Coulomb interactions, the sampling is subject to a phase problem. The phase problem can be remedied approximately by constraining the random walks using a trial wavefunction \cite{zhang_quantum_2003}, which is the only approximation in the method. The quality of the approximation can, in principle, be improved by improving the quality of the trial wavefunction. However, even with a simple Hartree-Fock-based trial wavefunction, AFQMC has been established as a high-precision many-electron method for molecules \cite{williams_direct_2020,shee-transformative-jcp-2023,motta_initio_2018,lee_twenty_2022}, solids \cite{purwanto_pressureinduced_2009,ma_quantum_2015,chen_initio_2021}, and Hubbard-like models \cite{leblanc_solutions_2015,qin_absence_2020,xiao_temperature_2023,xu_coexistence_2024}.
Our AFQMC calculations follow the QED extension of the method in the Coulomb gauge as described in \cite{weber_phaseless_2025}. The breaking of the long-wavelength approximation in \cref{eq:ham} leads to additional spatial dependencies of the electron-photon coupling via quadrupole and angular momentum matrix elements. However, this presents no fundamental challenge for the method. Our trial wavefunctions are the UHF-coherent-state product states obtained using the method above. We use 2000 walkers and an imaginary time step of $\Delta\tau = 0.005~\text{Ha}^{-1}$, performing reorthogonalization every 10, measurements every 10, and population control every 5 timesteps, respectively. The simulation code uses the Carlo.jl~\cite{weber_carlojl_2025} framework for parallel scheduling and statistical postprocessing.
\subsection{QED-FCI}
Our FCI calculations for H and H\textsubscript{2} work in a product basis of electronic Fock states and photonic occupation number states, subject to the occupation number cutoff $n_\text{ph} < 40$, which we find well converged. The diagonalization is performed using KrylovKit.jl~\cite{Haegeman_KrylovKit_2024}.

\section{Data availability}
All simulation data, along with the code used to generate the figures, is available publicly in Ref.~\cite{dataset}.
\bibliography{paper.bib}
\section{Acknowledgements}
We thank Florian Kluibenschedl for inspiring discussions about electronic cavities coupled to quantum rotors. The Flatiron Institute is a division of the Simons Foundation.
\section{Competing interests}
The authors declare no competing interests.
\clearpage

\appendix
\section*{Supplementary Note 1}
\setcounter{figure}{0}
\renewcommand{\thefigure}{S\arabic{figure}}

\subsection{Origin dependence}
In this section, we first discuss the intrinsic lack of translation symmetry in our model, propose an alternative few-mode model that retains complete translation symmetry, and finally comment on additional, artificial sources of origin-dependence in our finite basis-set calculations.

\begin{figure}[h]
\includegraphics{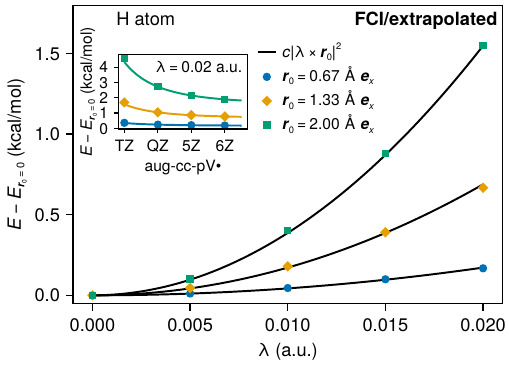}
\caption{\textbf{Intrinsic origin dependence due to quantum electric field.} The energy difference between an H atom at $\vb r_0$ and an H atom at the mode origin. The energies are extrapolated to the basis-set limit using a cubic extrapolant, $E(n) = E(\infty) + a_{\lambda,\vb r_0}/n^3$, where $n$ is the basis set order of the aug-cc-pVQZ, 5Z, and 6Z basis sets (inset). The resulting energy differences show good agreement with the predicted quadratic scaling $|\vlambda \times \mathbf{r}_0|^2$ with a proportionality factor $c \approx 1.54\,\text{Ha}^{-1}$.}
\label{fig:h_origin}
\end{figure}

At first glance, the Pauli-Fierz Hamiltonian in \cref{eq:pfierz} with a single-mode field $\vu{A}(\vb r) = \Qph \vlambda \times \vb{r}/\sqrt{\Omega}$ resembles a translation invariant model, where the choice of origin is arbitrary up to a gauge transformation. For a classical field, $\Qph = \text{const}$, this is indeed the case, as can be seen by applying the unitary transform $U = e^{i \Qph (\vlambda \times \vb{r}_0)\cdot \vb r/\sqrt{\Omega}}$. For a quantum field, however, this transformation does not commute with the photon momentum $\Pph$ so that a change in the origin can no longer be gauged away.

This origin dependence has a physical reason. If quantum effects are included, the uniform magnetic field does not exist on its own. It coexists with electrical field fluctuations, which, for our single-mode cavity, take the shape of a vortex field around the origin (proportional to $\vlambda \times \vb{r}$). The mode origin where $\vu A (\vb r)$ vanishes is therefore no longer an arbitrary choice but the physical point in the cavity where the electric field fluctuations vanish.

In our calculations, we assume that the molecules are very close to this mode origin. This may not be true for a realistic ensemble of molecules, and in the following, we will quantify the effect of the origin dependence in such situations. First, we note that at the Hartree-Fock level, there is no intrinsic origin dependence at all. Because we are effectively treating $\Qph$ as a classical mean-field, we can shift the origin using regular gauge transformations. Consequently, only the light-matter correlation energy is origin dependent, which, in turn, means that the origin dependence is a higher-order process perturbatively scaling with at least $\mathcal{O}(|\vlambda \times \vb r_0|^2)$, for a displacement $\vb r_0$ from the mode origin perpendicular to $\vlambda$. We confirm this scaling numerically in \cref{fig:h_origin}. If $|\vb r_0|$ is large compared to the size of the molecule, the vortex E field can be approximated as a uniform field and should lead to similar effects to those that have been widely studied in the literature for electric cavities in the dipole approximation. At even larger $|\vb r_0|$, the electric field will grow stronger indefinitely, and our model (which assumes a spatially constant $B$ field everywhere in space) becomes unrealistic.

For future studies of magnetic cavities, we propose two approaches to address this issue in a more controlled manner. The first path is to embrace the origin dependence and start from \emph{ab initio} parameters of a specific magnetic cavity. The second path is to restore complete translation invariance by adding at least one additional degenerate-frequency mode, e.g.,
\begin{align}
\hat{\mathbf{A}}(\mathbf{r}) = \veps \cos(\mathbf{q} \cdot \mathbf{r}) Q_+ + \veps \sin(\mathbf{q} \cdot \mathbf{r}) Q_-. \label{eq:newA}
\end{align}
Such pairs of modes would arise naturally in a cavity with periodic boundary conditions in the direction of $\mathbf q$.

We note that the vector potential in \cref{eq:newA} is translation invariant under $\mathbf{r}\mapsto \mathbf{r} + \mathbf{r}_0$, combined with a rotation of the vector $(Q_+, Q_-)$ by an angle $\mathbf{q} \cdot \mathbf{r}_0$. Since $Q_+$ and $Q_-$ are posed to have the same frequency, the photon kinetic term is also invariant under such rotations and the Pauli-Fierz Hamiltonian with the given multi-mode vector potential would be translation invariant. A similar discussion can be made for rotation invariance, which would require including additional pairs of symmetry-related modes.

Finally, we discuss the issue of artificial origin dependence due to basis set truncation. From quantum-chemical calculations with classical magnetic fields, it is well known that gauge transformations are exact only in the complete-basis-set limit. This can lead to a spurious nonintrinsic origin dependence, where the chosen basis-set truncation can represent the magnetic states well for one choice of origin but not for another. To work around this problem, London-type orbitals have been developed~\cite{london_theorie_1937,ditchfield_selfconsistent_1974} that include an exact gauge transformation in their definition. Adapting these techniques for quantum fields, however, leads to interaction integrals that depend explicitly on $\Qph$ and would therefore need to be recomputed throughout the computation at a prohibitive cost. Therefore, in this work, we mitigate the nonintrinsic origin dependence in a brute-force way by simulating in large enough standard Gaussian basis sets.

\begin{figure}
\includegraphics{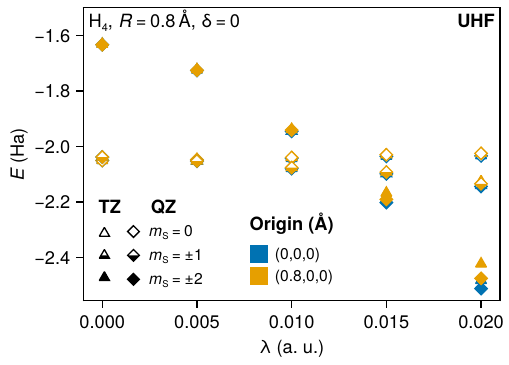}
\caption{\textbf{Nonintrinsic origin dependence due to basis-set error.} The total energy of \Hn{4} as a function of cavity coupling $\lambda$ with and without origin shift. Calculations are performed at zero dimerization $\delta$ and fixed radius $R$ for the aug-cc-pVTZ (TZ) and aug-cc-pVQZ (QZ) basis sets.}
\label{fig:origin}
\end{figure}
\Cref{fig:origin} shows UHF results for \Hn{4} for both unshifted and shifted origin. Even though Hartree-Fock is expected not to pick up the intrinsic origin dependence, we still observe nonintrinsic origin dependence. This error decreases in the larger basis set and is negligible for both aug-cc-pVTZ and aug-cc-pVQZ basis sets in the region $\lambda < 0.01$ a.u., where we find the geometry transition. We also find that the energies without shifted origin are strictly lower than the ones with, indicating that centering the molecule at the origin minimizes the basis set error.

\subsection{Basis set dependence at large $\lambda$}
\begin{figure}
\includegraphics{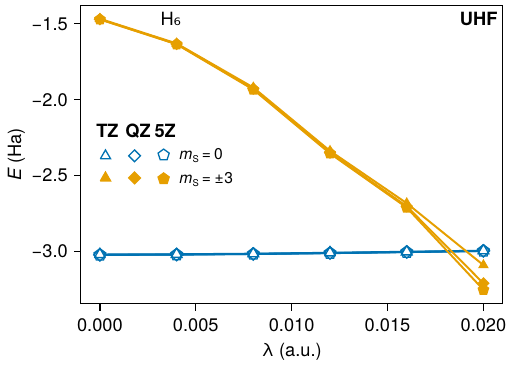}
\caption{\textbf{Basis set dependence for very strong coupling.} Shown is the QED-UHF energy of the undistorted \Hn{6} ring with $R=0.7$~\AA for different magnetizations $\ms$ as a function of the light-matter coupling $\lambda$.}
\label{fig:basis_set}
\end{figure}

Apart from the origin dependence issue, there is also a general challenge of converging to the complete basis set limit if the light-matter coupling is large enough and the ground state is strongly magnetized. We will illustrate this effect using the \Hn{6} ring from the main text as an example. \Hn{6} is not distorted outside of the cavity and therefore does not undergo a symmetrizing geometry transition under light-matter coupling. Yet, at very strong light-matter coupling, it shows signs of a fully polarized ground state (\cref{fig:basis_set}). A similar feature is also present in the other H rings. It is important to note, however, that the convergence within conventional Gaussian basis sets is very slow in this regime, likely due to difficulties of resolving the angular momentum coupling operator within these basis sets. For this reason, we did not include this regime in our analysis in the main text. Remarkably, cyclobutadiene does not display this instability towards a fully magnetized state for the couplings we studied. This could be attributed to the stronger diamagnetic contribution, visible in main text Fig. 4, which keeps strongly polarized states energetically separated from the ground state.

For classical magnetic fields, the basis set dependence issue is conventionally circumvented by using London-type orbitals~\cite{london_theorie_1937}. Adopting this approach for cavity fields is not straight-forward, because the cavity magnetic field fluctuates. In QED-AFQMC, this would necessitate recomputing the electron interaction tensor on the fly (or finding an efficient representation for its field-dependent form). An alternative approach could be to supplement the basis set with additional basis functions that increase the resolution of the magnetic coupling without increasing cardinality~\cite{sugimoto_gaugeinvariant_1995a}.

\end{document}